\documentstyle[12pt,epsf]{article}
\catcode`\@=11
\def\section{\@startsection {section}{1}{\z@}{3.5ex plus 1ex minus 
    .2ex}{2.3ex plus .2ex}{\Large\bf}}
\catcode`\@=12

\newcommand{\beq}{\begin{equation}}
\newcommand{\eeq}{\end{equation}}
\newcommand{\ben}{\begin{eqnarray}}
\newcommand{\een}{\end{eqnarray}}
\newcommand{\etc}{{\sl etc}}
\begin{document}
\begin{titlepage}
\vskip-30mm
\begin{flushright}
{\normalsize KANAZAWA95-19}
\end{flushright}
\vskip20mm 
\begin{center}
{\Large\bf The effectiveness of the local potential approximation\\
in the Wegner-Houghton renormalization group}
\end{center}
\vskip20mm
\begin{center}
Ken-Ichi Aoki, Kei-ichi Morikawa, Wataru Souma, Jun-ichi Sumi 
and Haruhiko Terao
\end{center}
\vskip10mm
\begin{center}
Department of Physics, Kanazawa University,
\end{center}
\begin{center}
Kakuma-machi, Kanazawa 920-11, Japan
\end{center}

\vfill
\begin{center} 
\begin{bf}
Abstract
\end{bf}
\end{center}
The non-perturbative Wegner-Houghton renormalization group  is analyzed
by the local potential approximation
in O($N$) scalar theories in $d$-dimensions $(3\leq d\leq 4)$.
The leading critical exponents $\nu$ are calculated in order to investigate
the effectiveness of the local potential approximation
by comparing them with the other non-perturbative methods.
We show analytically that 
the local potential approximation gives the exact exponents up to
$O(\epsilon )$ in $\epsilon$-expansion and
the leading in $1/N$-expansion.
We claim that this approximation 
offers fairly
accurate results 
in the whole range of the parameter space of $N$ and $d$.
It is a great advantage of our method that no diverging expansions 
appear in the procedure.

\vfill
\end{titlepage}
\eject
\section{Introduction}

The non-perturbative phenomena of the quantum field theories have been
fascinating many physicists.
There are only a limited number of tools to attack such problems, for example,
the Monte Carlo simulations in the lattice field theories,
the Schwinger-Dyson equations, $\epsilon$-expansion,
$1/N$-expansion \etc .
In this article we focus our attention on the Wilson 
renormalization group (RG) among these approaches\cite{WK}.
The Wilson RG equations are given by the form of the functional differential
equations for the so-called Wilsonian effective action defined in the
Euclidean space.
Therefore it is inevitable to approximate them for the practical
calculations.
We usually expand the effective action in terms of the number of 
derivatives included in the general operators and 
solve the Wilson RG equation within 
the subspace up to some finite number of
derivatives. 
In the first order of this approximation, any derivative couplings are dropped
except for the kinetic terms. This is called the local potential approximation
(LPA).
Then the functional differential equation for the effective action is reduced
to a non-linear partial differential equation for the local potential.
Instead of analyzing this partial differential equation directly
we may also expand the local potential with respect to the
fields and truncate the series at some finite order.
By this approximation we may solve the coupled differential
equations for the expansion coefficients, that is, the coupling constants.
These equations are much easier to solve than the original partial
differential equation.
Actually this series expansion is found to converge remarkably fast.

It is important to see whether the LPA
offers us sufficiently good results before performing the higher order
calculations in the derivative expansion.
For this purpose we take the Wegner-Houghton (W-H)
equation\cite{WH} among several formulations\cite{P,W} and compare the LPA
of this with $\epsilon$-expansion as well as
$1/N$-expansion in O($N$) scalar field theories.

The contents of this article are the following.
In section 2 we briefly review the W-H equation for O($N$) scalar
field theories and the LPA.
We series expand the potential and
calculate the correlation length critical exponent.
We compare our results with the  $\epsilon$-expansion in section 3 
and the $1/N$-expansion in section 4.
The accuracy of the LPA is discussed in comparison with the Borel
resummation of the $\epsilon$-expansion and other non-perturbative 
calculations.
We also argue the property of the LPA
in relation to the $1/N$-expansion.
The section 5 is devoted to the summary and discussions.

\section{The Wegner-Houghton equation and the local potential approximation}

First we derive the W-H equations for the O($N$) scalar theories\cite{WH}.
The starting point is the Euclidean path integral defined 
with momentum cutoff
$\Lambda(t)=e^{-t}\Lambda$;
\beq
Z=\int{\cal D}{\phi}\exp\left(- S_{\rm eff}[\Lambda(t)]\right)\ ,
\eeq
where $S_{\rm eff}$ is the Wilsonian effective action.
The RG differential equations give the response of $S_{\rm eff}$
under the infinitesimal change of the cutoff $\Lambda$ with keeping
the partition function $Z$ unaltered. 
The difference in $S_{\rm eff}$ induced by the change of the cutoff is 
determined by integrating the ``shell mode'' with the momenta
between $\Lambda(t)$ and $\Lambda(t+\delta t)$.
It should be noted that this integration is reduced to the Gaussian one
for infinitesimally small $\delta t$, and is exactly carried out.
Besides we rescale the momentum and the fields by $\Lambda$,
since the change of the dimensionless quantities are of our interest.
Thus we deduce the W-H equation for O($N$) scalar field theories as
\ben
\frac{dS_{\rm eff}}{dt}&=&
\frac{1}{2{\delta}t}{\int_{e^{-{\delta}t}\leq|p|\leq1}}\frac{d^{d}p}{(2\pi)^d}
\left({\rm ln}
\frac{{\delta}^{2}S_{\rm eff}}{\delta\phi(p)
\delta\phi(-p)}
\right)_{\alpha\alpha}
\nonumber
\\
&-&
\frac{1}{2{\delta}t}\int
\int_{e^{-{\delta}t}\leq|p|,|q|\leq 1}
\frac{d^dp}{(2\pi)^d}
\frac{d^dq}{(2\pi)^d}
\left\{
\frac{{\delta}S_{\rm eff}}{\delta\phi^\alpha(p)}
\left(
\frac{{\delta}^{2}S_{\rm eff}}{\delta\phi(p)\delta\phi(q)}
\right)^{-1}_{\alpha\beta}
\frac{{\delta}S_{\rm eff}}{\delta\phi^\beta(q)}
\right\}
\nonumber
\\
&+&
\int_{|p|\leq1}\frac{d^{d}p}{(2\pi)^d}{\phi}^\alpha(p)
\left(
\frac{2-\eta-d}{2}-p^{\mu}\frac{{\partial}'}{{\partial}p^\mu}
\right)
\frac{\delta}{\delta\phi^\alpha(p)}S_{\rm eff}
+d{\cdot}S_{\rm eff}. \label{eq:WHON}
\een
Here $\phi^\alpha$ denote $N$-component scalar fields and
$\eta / 2$ is the anomalous dimension of these fields.

The effective action $S_{\rm eff}$ is given in the derivative expansion by
\beq
S_{\rm eff}=\int d^dx\left\{
V(\rho)+\frac{1}{2}Z_1(\rho)(\partial_\mu\phi^\alpha)^2+
\frac{1}{2}Z_2(\rho)\phi^\alpha\partial^2\phi^\alpha+.....\right\},
\eeq
where we introduced the O($N$) invariant operator 
$\rho=\frac{1}{2}\phi^\alpha\phi^\alpha$.
If we substitute this into Eq.(\ref{eq:WHON}), the W-H equation
turns
out to be the coupled partial differential
equations for $V(\rho)$, $Z_i(\rho)$ $(i=1,2,...)$.
In the LPA we restrict the $S_{\rm eff}$ to
\beq
S_{\rm eff}=\int d^dx\left\{
V(\rho)+\frac{1}{2}(\partial_\mu\phi^\alpha)^2
\right\}.
\eeq
Note that the kinetic term is not renormalized in the LPA,
which means that the anomalous dimension automatically vanishes.
Then the W-H equation in the LPA is written down as
\beq
\frac{dV}{dt}=\frac{A_d}{2}\left[{\rm ln}(1+V^\prime+2{\rho}V^{\prime\prime})
+(N-1){\rm ln}(1+V^\prime)\right]
+d{\cdot}V+(2-d){\rho}V', \label{eq:LPAWHON}
\eeq
where the prime denotes the derivative with respect to $\rho$ and $A_d$ is
the $d$ dimensional angular integral,
$A_d=\pi^{-d/2}2^{1-d}/\Gamma(d/2)$. 

Next we are going to examine this partial differential equation by 
expanding the potential into a power series
and by truncating it at some order.
Such analysis based on the Taylor series will be useful when comparing
the LPA with the $\epsilon$-expansion and 
also the $1/N$-expansion as we will see later.
If the truncation method gives the same results 
irrespective to the order of the truncation,
we consider that they are the LPA results.

The naive expansion may be to expand around the origin (Fixed Scheme),
\beq
V(\rho)=\sum_{m=0}^{M}\frac{a_m(t)}{m!}\rho^m. \label{eq:ORIGIN}
\eeq
Also it will be natural to expand around the minimum of the potential
$b_1(t)$ (Comoving Scheme)\cite{A},
\beq
V(\rho)=b_0(t) + 
\sum_{m=2}^{M}\frac{b_m(t)}{m!}(\rho-b_{1}(t))^m .\label{eq:MINI}
\eeq
These two schemes give the RG flows projected onto the different
$M$-dimensional subspaces. In Fig.1, we show the leading critical
exponents $\nu$ (the correlation length exponents) estimated by these
two truncation schemes for $N$=1 theory in $d$=3. It is seen
that the Comoving Scheme converges very rapidly to a value 
0.68956(1), which should be compared with 0.687(1) obtained 
by the analysis of the partial differential equation (5)\cite{HH}.
Contrary to this good convergence, the exponents estimated by the
Fixed Scheme seem to remain oscillating. The reason of this oscillating
behavior has been discussed by Morris\cite{M1}. Anyhow this convergence
indicates that the relevant operator can be described sufficiently well
within quite a few dimensional subspace given by Eq.(7)\cite{AMSST}.

\begin{figure}[htb]
\epsfxsize=0.75\textwidth
\centerline{\epsffile{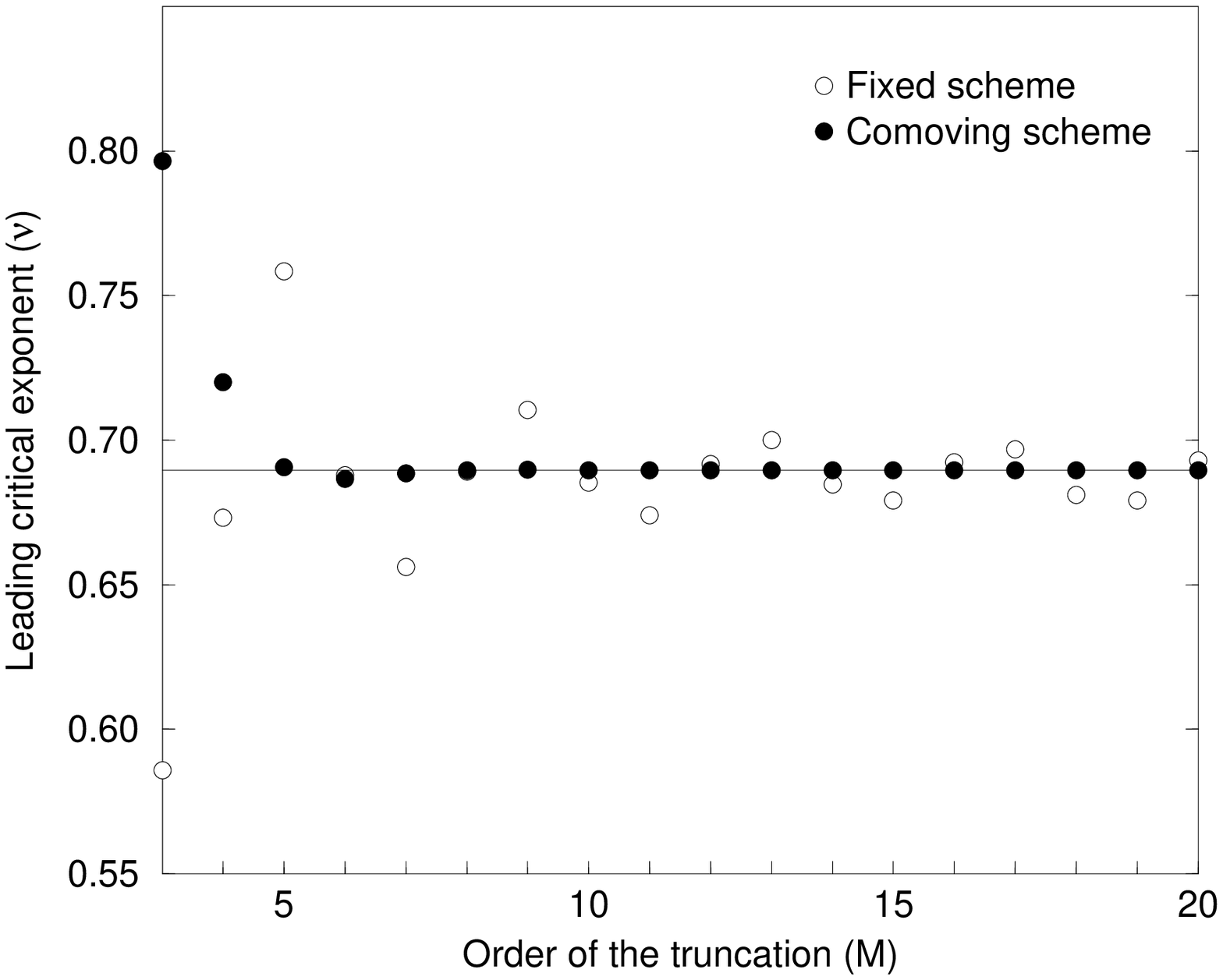}}
\parbox{6.5in}{
\footnotesize 
Figure 1:
The critical exponents for the $N$=1 scalar theory in $d$=3
calculated by the Fixed Scheme and the Comoving Scheme 
as a function of the order 
$M$ of the truncation. 
}
\end{figure}

Now let us consider general coordinates introduced to describe the RG flows.
Suppose $\cal{M}$ is the theory space of infinite dimension and $\{a^i\}$
is a generic coordinate system (a set of coupling constants) of
$\cal{M}$.
The operator $d/dt$ defines a vector field in $\cal{M}$ and is given in the
coordinate system $\{a^i\}$ by
\beq
\frac{d}{dt}=\beta^i(a)\frac{\partial}{\partial a^i},
\eeq
where we define the generalized beta functions $\beta^i(a)=da^i/dt$.
This means that the beta functions are the components of the 
vector field $d/dt$.
For any function $S(a^i)$ defined in $\cal{M}$, $dS/dt$ gives a tangent
vector in the space of $\{S\}$;
\beq
\frac{dS}{dt}=\beta^i(a)\xi_i(a),
\eeq
where we also introduced the base vectors in this space, $\xi_i=\partial S/
\partial a^i$.
In practice the RG equation gives us $dS/dt$, where $S$ stands for the
Wilsonian effective action, and we may obtain the beta functions by expanding
$dS/dt$ by the base vectors $\xi_i$.
Here it should be noted that the base vectors $\xi_i$ depend on $a^i$,
that is, the position in $\cal{M}$ generically. We may call such coordinates
system a ``comoving frame''.
The expansion at the potential minimum given in Eq.(\ref{eq:MINI}) is a
typical example of
this comoving frame, since the base vectors $\xi_i$ given explicitly by 
$\xi_1=-b_2-b_3(\rho-b_1)-\cdot\cdot\cdot$, 
$\xi_2=\rho-b_1$, 
$\xi_3=(\rho-b_1)^2$, 
depend on the coordinates $b_i$ indeed.

Suppose we may choose the coordinates $\{a^i\}$ so that the matrix $\Omega^i_j=
\partial\beta^i/\partial a_j$ forms a lower triangle,
\beq
\Omega^i_j(a)=\frac{\partial\beta^i}{\partial a^j}(a)=
0\;\;\;{\rm for}\;\;\;i<j ,
\eeq 
then the components $a^i(t)$ of the flow (curve) can be exactly evaluated
within the finite dimensional subspaces of $\cal{M}$ spanned by 
$(a^1, a^2, \cdots a^i)$.
Here let us call such a special coordinate system the ``perfect coordinates''.
Generally it would be difficult to find such coordinates as much as to
solve the RG equation exactly.
However in the large $N$ limit the expansion around the potential minimum will
be found to give us an example of the ``perfect coordinates''.
Actually we can solve the RG equation in large $N$ limit 
exactly in the every order of the truncation.

\section{The comparison with the $\epsilon$-expansion.}

The critical exponents have been calculated in powers of
$\epsilon=4-d$
by Wilson and Fisher\cite{WF,WK}. The correlation length exponent $\nu$
is found to be
\beq
\nu_{\epsilon}=\frac{1}{2}+\frac{N+2}{4(N+8)}\epsilon
+\frac{(N+2)(N^2+23N+60)}{8(N+8)^3}\epsilon^2+O(\epsilon^3). \label{EPS}
\eeq
In this section we consider to compare the exponent obtained by the
W-H equation in the LPA with this result.

We examine Eq.(5) in $d=4-\epsilon$.
In the Fixed Scheme defined by Eq.(6) the first three beta functions are given by
\ben
\beta_1&=&\frac{A_{4-\epsilon}}{2}\left[\frac{3a_2}{1+a_1}+
(N-1)\frac{a_2}{1+a_1}\right]+2a_1 , \nonumber \\
\beta_2&=&\frac{A_{4-\epsilon}}{2}\left[\frac{5a_3}{1+a_1}-
\frac{9a_2^2}{(1+a_1)^2}+(N-1)\left\{\frac{a_3}{1+a_1}-\frac{a_2^2}{(1+a_1)^2}
\right\}\right]+\epsilon a_2 , \label{eq:MINIBETA} \\
\beta_3&=&\frac{A_{4-\epsilon}}{2}\left[\frac{7a_4}{1+a_1}
-\frac{45a_2a_3}{(1+a_1)^2}+\frac{54a_2^3}{(1+a_1)^3}
\right. \nonumber \\
& &\;\;\;\;\;\left.+(N-1)\left\{\frac{a_4}{1+a_1}-\frac{3a_2a_3}{(1+a_1)^2}
+\frac{2a_2^3}{(1+a_1)^3}\right\}\right]
+2(\epsilon-1)a_3 \nonumber. 
\een
From these equations the non-trivial fixed point solution is found to be
\ben
a_1^*&=&-\frac{N+2}{2(N+8)}\epsilon-\left\{\frac{(N+2)^2}{4(N+8)^2}+
\frac{(N+2)(20N+88)}{2(N+8)^3}\right\}\epsilon^2 +O(\epsilon^3), \nonumber \\
a_2^*&=&\frac{2}{A_4(N+8)^3}\epsilon+
\frac{\left\{A_4(20N+88)-A_4'(N+8)^2\right\}}{A_4^2(N+8)^3}\epsilon^2 
+O(\epsilon^3), \\
a_3^*&=&\frac{4(N+26)}{A_4^2(N+8)^3}\epsilon^3+O(\epsilon^4) , \nonumber
\een
in powers of $\epsilon$, where
$A_4=1/8\pi^2$,  $A_4'=(1-\gamma_E+{\rm ln}4\pi)/16\pi^2$ and
$\gamma_E=0.5772....$ denotes the Euler constant.
The exponent $\nu$ is given by the positive eigenvalue of the
matrix
$\Omega^i_j=\partial\beta_i/\partial a_j$
at the non-trivial fixed point.
Thus the exponent estimated in the LPA turns out to be
\beq
\nu=\frac{1}{2}+\frac{N+2}{4(N+8)}\epsilon
+\frac{(N+2)(N^2+38N+96)}{8(N+8)^3}\epsilon^2+O(\epsilon^3).
\eeq
Comparing with Eq.(11) the LPA exponent is exact in 
$O(\epsilon)$\cite{WH}.
The deviation starts in $O(\epsilon^2)$, which is due to the
ignorance of the renormalization of the derivative couplings.

\begin{table}
\begin{center}
\begin{tabular}{|c|c|c|c|c|c|c|c|c|c|c|c|}\hline
$d$&4.0&3.9&3.8&3.7&3.6&3.5&3.4&3.3&3.2&3.1&3.0\\\hline
$\nu$&0.500&0.509&0.519&0.531&0.544&0.560&0.577&0.598&0.622&0.652&0.6896\\\hline
\end{tabular}
\end{center}
\parbox{6.5in}{\begin{center}
\footnotesize Table 1: The critical exponents for the $N$=1 scalar
theory in $d$-dimension.
\end{center}
}
\end{table}

\begin{figure}
\epsfxsize=0.75\textwidth
\centerline{\epsffile{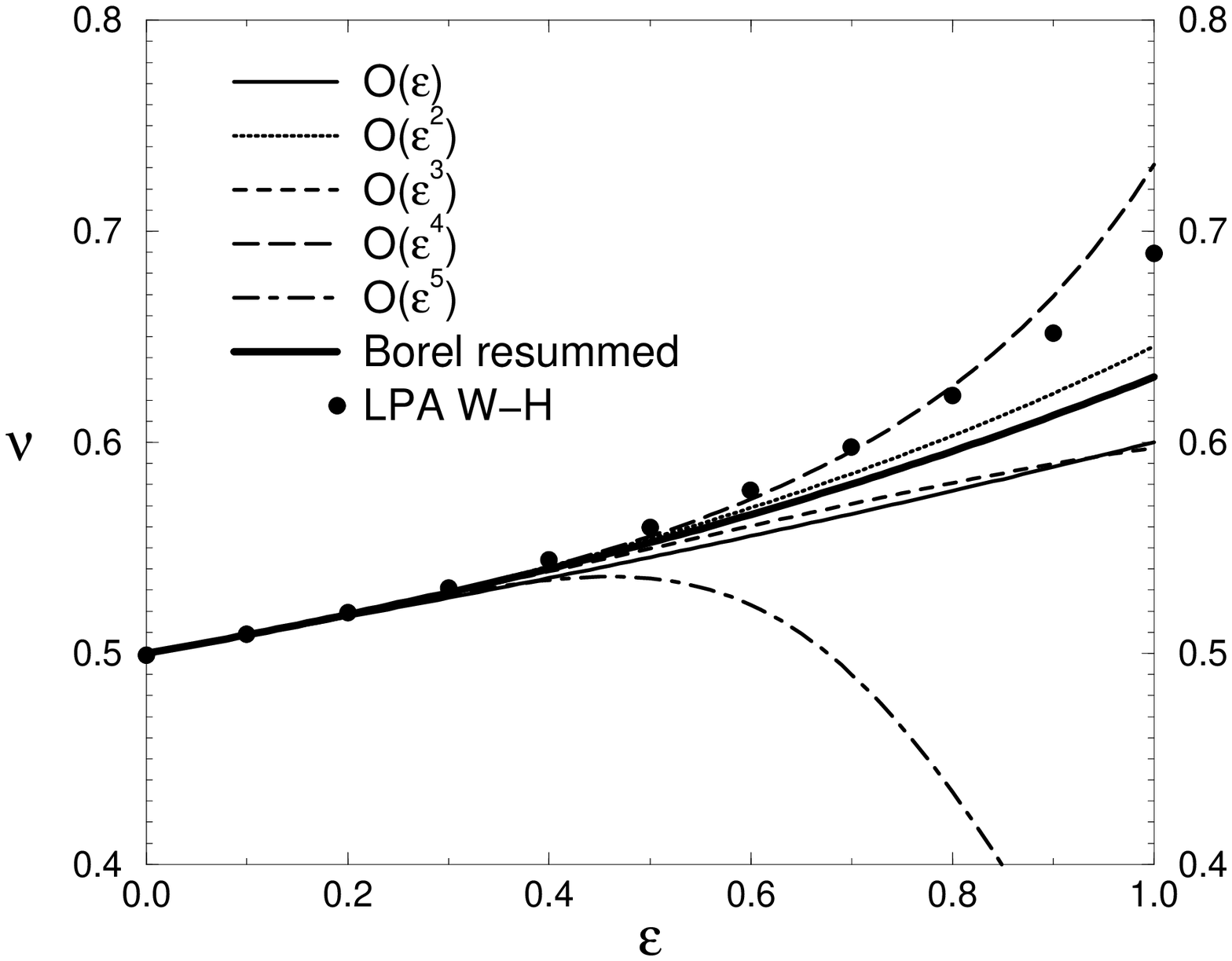}}
\parbox{6.5in}{
\footnotesize Figure 2:
The critical exponents for the $N$=1 scalar theory as a function of $\epsilon$.
The estimations by the W-H equation in the LPA and by
$\epsilon$-expansion up to $O(\epsilon^5)$ are shown. 
The bold solid line is by the Borel resummation of the $\epsilon$-expansion.
}
\end{figure}

We list the exponents calculated by the W-H equation in Table 1, and
show them in Fig.2  as well as the exponents obtained 
by $\epsilon$-expansion for $N$=1\cite{VKT,KNSCL}.
The $\epsilon$-expansion itself produces an
asymptotic series, which means the series converges up to a certain
order but eventually turns to diverge.
Therefore the $\epsilon$-expansion would not be meaningful,
unless this asymptotic series is summed up to a finite value.
Actually not only the $\epsilon$-expansion but also the 
ordinary perturbation with respect to the coupling constants 
has been known to generate an asymptotic series. 
Since the late 70's the Borel summation technique has been found
to be quite powerful to sum up such asymptotic series and has
succeeded in offering the reliable values for the critical exponents 
\cite{L,VKT,GZ}.
The critical exponent $\nu$ calculated by the
Borel summation technique {\sl \`{a} la} \cite{VKT,GZ} as a function of 
$\epsilon$ is also drawn in Fig.2. 
This may well be regarded as the accurate result for the
exponent with which we can compare our RG calculations.
Then our method gives comparatively
good values for the exponent, if we take into consideration that the
LPA is just the first order approximation in the derivative expansion. 

Of course it will be extremely important to investigate whether the
Wilson RG calculation in the higher order approximation than the LPA
really gives the results converging order by order.
However it should be noted that 
the Wilson RG seems to give much easier methods to handle than
the $\epsilon$-expansion and also than the perturbative expansion
at least in order to 
obtain moderately reliable results.
Because the higher order of 
such expansions necessitates quite
complicated calculations. Moreover it is essential to know 
the large order behavior of the asymptotic series in performing the 
resummation. Besides it would be necessary to tune some
free parameters introduced to define the Borel transformation
so as to reach the accuracy shown in Ref.\cite{VKT,GZ}.
Thus the procedure of the Borel summation is rather complicated.
On the other hand
no such expansions leading to asymptotic series are
carried out in our calculation of the Wilson RG equation. Therefore
we may say that our RG method offers directly the results
which are supposed to be sums of asymptotic series appearing
in other methods. Thus this
is one of the characteristic and advantageous features of the
Wilson RG approach.  

\section{The comparison with the $1/N$-expansion.}

It has been recognized\cite{WH,RTW} that the
LPA for the $N$-vector model gives the exact effective potential
in large $N$ limit. To see this, 
we expand the effective action as $S_{\rm eff}=V_{\rm eff}
+S_{\rm eff}^2
+S_{\rm eff}^3+....$, where the index denotes the number of the 
operator $\phi^2$ with non-vanishing momentum.The beta-functional
of $V_{\rm eff}$ depends only on  $\{V_{\rm eff}\}$ 
in large $N$ limit\cite{WH}.
That is, $V_{\rm eff}$ can be seen as a sort of the first
component of the perfect coordinates introduced in Eq.(10).
Therefore we may solve the RG equation exactly in the subspace
irrespectively to other subspaces. The RG
equation for this subspace turns out to be the W-H
equation (\ref{eq:LPAWHON}).
This feature is also expected from the fact that
the renormalization of the derivative couplings is required
only in $O(1/N)$\cite{S}, which is discarded in the LPA.

Here we also show 
this remarkable property by deducing the RG equation directly
from the large $N$ effective potential\cite{RTW}.
In order to define the large $N$ limit RG equation,
we introduce $x\equiv\rho/N$ and 
$F(x,t) \equiv \partial V(\rho)/\partial \rho$, and rewrite
Eq.(5) as 
\beq
\frac{dF}{dt}(x,t)=\frac{A_{d}}{2N}
\left\{
\frac{3F'+xF''}{1+2F+xF'} + (N-1)\frac{F'}{1+F}
\right\}
+2F+(2-d)xF'.
\eeq
In large $N$ limit this equation is reduced to
\beq
\frac{dF}{dt}(x,t)=\frac{A_{d}}{2}
\frac{F'}{1+F} +2F+(2-d)xF'. \label{eq:LARGEN}
\eeq

Now we are going to show this equation can be derived directly from
the gap-equation in large $N$ limit as well. First introduce
two auxiliary fields $\chi$ and $\rho$,
\beq
Z= \int {\cal D}\phi^\alpha {\cal D}\chi {\cal D}\rho
\exp \left\{
-\int d^d x \left[ \frac{1}{2}(\partial\phi^\alpha)^2 
+ \chi\left( {1\over 2}\left(\phi^\alpha\right)^2  - N \rho \right)
+N V(\rho)
\right]\right\},
\eeq
so that we may treat the generic
form of the potential. Then the effective potential $V_{\rm eff}$,
to say strictly, the constraint effective potential\cite{FK}, 
for $N$-vector model may be written as
\beq
\exp\left(-N V_{\rm eff}(\bar{\rho})\right)
= \int {\cal D}\chi {\cal D}\rho
\exp \left\{
-N \int d^d x \left[\chi \left(\bar{\rho} - \rho\right)
+V(\rho)\right] -\frac{N}{2} {\rm trln}(-\partial^2 + \chi)
\right\},
\eeq
where $\bar{\rho}$ denotes $\bar{\phi}^2/2N$.
In large $N$ limit, the path integral of these auxiliary fields is
evaluated by the saddle point method.  
Then the effective potential $V_{\rm eff}$ will be given by solving 
the following coupled equations;
\begin{eqnarray}
V_{\rm eff}[\bar{\rho}] &
=&\chi (\bar{\rho} - \rho) + V(\rho)
+ \frac{1}{2}\int_{\Lambda} \frac{d^dk}{(2 \pi)^d} \ln(k^2 + \chi),\\
\rho &=& 
\bar{\rho} + 
\frac{1}{2}\int_{\Lambda} \frac{d^dk}{(2\pi)^d} \frac{1}{k^2 +\chi},\\
\chi &=&
\frac{dV(\rho)}{d\rho} .
\end{eqnarray}
Here the cut-off $\Lambda$ in the momentum integration 
is introduced to derive the RG equation for the effective 
potential $V_{\rm eff}$. 
By considering the
infinitesimal change of the cut-off, the RG equation is
found to be
\beq
\frac{dV_{\rm eff}}{d \Lambda}
= \frac{\partial V_{\rm eff}}{\partial \Lambda}
= -\frac{A_{d}}{2} \Lambda^{d-1}\ln(\Lambda^2 + \chi),
\eeq
where the saddle point equations 
$\partial V_{\rm eff}/\partial \rho =0$ and
$\partial V_{\rm eff}/\partial \chi =0$ are used. 
Noting that $\chi$ is
given in terms of $V_{\rm eff}$ as $\chi = dV_{\rm eff}/d\bar{\rho}$, 
we eventually
obtain the RG equation for the effective potential $V_{\rm eff}$ as
\beq
\frac{dV_{\rm eff}}{d \Lambda}[\bar{\rho};\Lambda]
= -\frac{A_{d}}{2} \Lambda^{d-1}\ln(\Lambda^2 + V_{\rm eff}'),
\eeq
which is found to be identical to the W-H equation in
the LPA (\ref{eq:LARGEN}), after rescaling of the variables.

Next we solve this 
RG equation by projecting it on much smaller subspaces.
Actually if we expand the effective potential in powers of
$\rho - b_{1}(t)$, where $b_{1}$ is the minimum of the 
potential, as is given by Eq.(\ref{eq:MINI}), 
then $\beta_{k} \equiv d b_{k}/dt$
($k=1, 2, 3,...$) are given by
\begin{eqnarray}
\beta_{1} &=&-\frac{A_d}{2}+(d-2)b_1 , \nonumber \\
\beta_{2} &=&-\frac{A_d}{2}b_2^2+(4-d)b_2 , \\
\beta_{3} &=&-\frac{A_d}{2}(3b_2b_3+2b^3_2)+(6-2d)b_3, \ \ 
{\rm \etc} . \nonumber
\end{eqnarray}
Here note that, contrary to Eq.(\ref{eq:MINIBETA}),
 $\beta_{k}$ depends only on the 
couplings $b_{i}$ of $i \leq k$.
Namely this ``comoving frame'' is found to be the ``perfect
coordinates'' in large $N$ limit. 
This means that the flow of $b_{i}$ can be exactly 
determined independently of the higher order couplings
$b_{k}$ of $k > i$.
In other words the projection to each order of truncated subspace
in the comoving frame is always exact. 
Therefore the coordinates of the fixed point are also exactly 
solved in each order of the truncation and are found to be
\begin{eqnarray}
b_{1}^* &=& \frac{A_{d}}{2(d-2)}, \nonumber \\
b_{2}^* &=& \frac{2(4-d)}{A_{d}},\\
b_{3}^* &=& \frac{4(4-d)^3}{A_{d}^2 (d-6)}, \ \ \etc .\nonumber
\end{eqnarray}
In this perfect coordinates the matrix  
$\Omega_{i}^{j} \equiv \partial \beta_{i}/\partial b_{j}$ 
is characterized by a lower triangular form. 
At the non-trivial fixed point, $\Omega$ is given by
\beq
\Omega=
\left(
\begin{array}{c}
\begin{array}{cccccc}
d-2&0&0&0&0&\cdot\cdot\cdot\cdot\\
0&d-4&0&0&0&\cdot\cdot\cdot\cdot\\
0&*&d-6&0&0&\cdot\cdot\cdot\cdot\\
0&*&*&d-8&0&\cdot\cdot\cdot\cdot
\end{array}
\\
\cdot\cdot\cdot\cdot\cdot\cdot\cdot\cdot\cdot\cdot\cdot\cdot\cdot\cdot\cdot
\cdot\cdot\cdot\cdot\cdot
\end{array}
\right),
\eeq
where $*$ denotes some functions of $d$.
The eigenvalues of this triangular matrix are nothing but the 
diagonal elements, $d-2m, (m=1,2,3,..)$. 
Therefore the leading critical exponent in large $N$ limit 
is exactly derived as
\beq
\nu = \frac{1}{d-2}. \label{eq:LARGENLIMIT}
\eeq

Now we study the $1/N$ dependence.
In $d$=3, the critical exponent $\nu$ has been 
calculated up to $O(1/N^2)$\cite{AOO}, 
\beq
\nu = 1-\frac{32}{3N\pi^2}+
\frac{32}{N^2\pi^4}\left(\frac{112}{27}-\pi^2\right)
+O(\frac{1}{N^3}).
\eeq
In Table 2, we list the critical exponents calculated by our
RG method, and show them in Fig.3 as well as those
obtained by $1/N$-expansion and by other methods.
The RG equation in the LPA (5) is fairly effective in the
whole range of $N$, while $1/N$-expansion results show up
its divergent nature and there has been no way to sum up the 
series.

\begin{table}[htb]
\begin{center}
\begin{tabular}{|c|c|c|c|c|c|c|c|c|c|}\hline
$N$&1&2&3&4&5&10&20&50&100\\ \hline
$\nu$&0.6896&0.767&0.826&0.865&0.891&0.946&0.974&0.990&0.995\\ \hline
\end{tabular}
\end{center}
\parbox{6.5in}{\centerline
{\footnotesize Table 2. The critical exponents of the O($N$)-symmetric 
scalar theories in $d$=3.
}}
\end{table}

\begin{figure}[p]
\epsfxsize=0.75\textwidth
\centerline{\epsffile{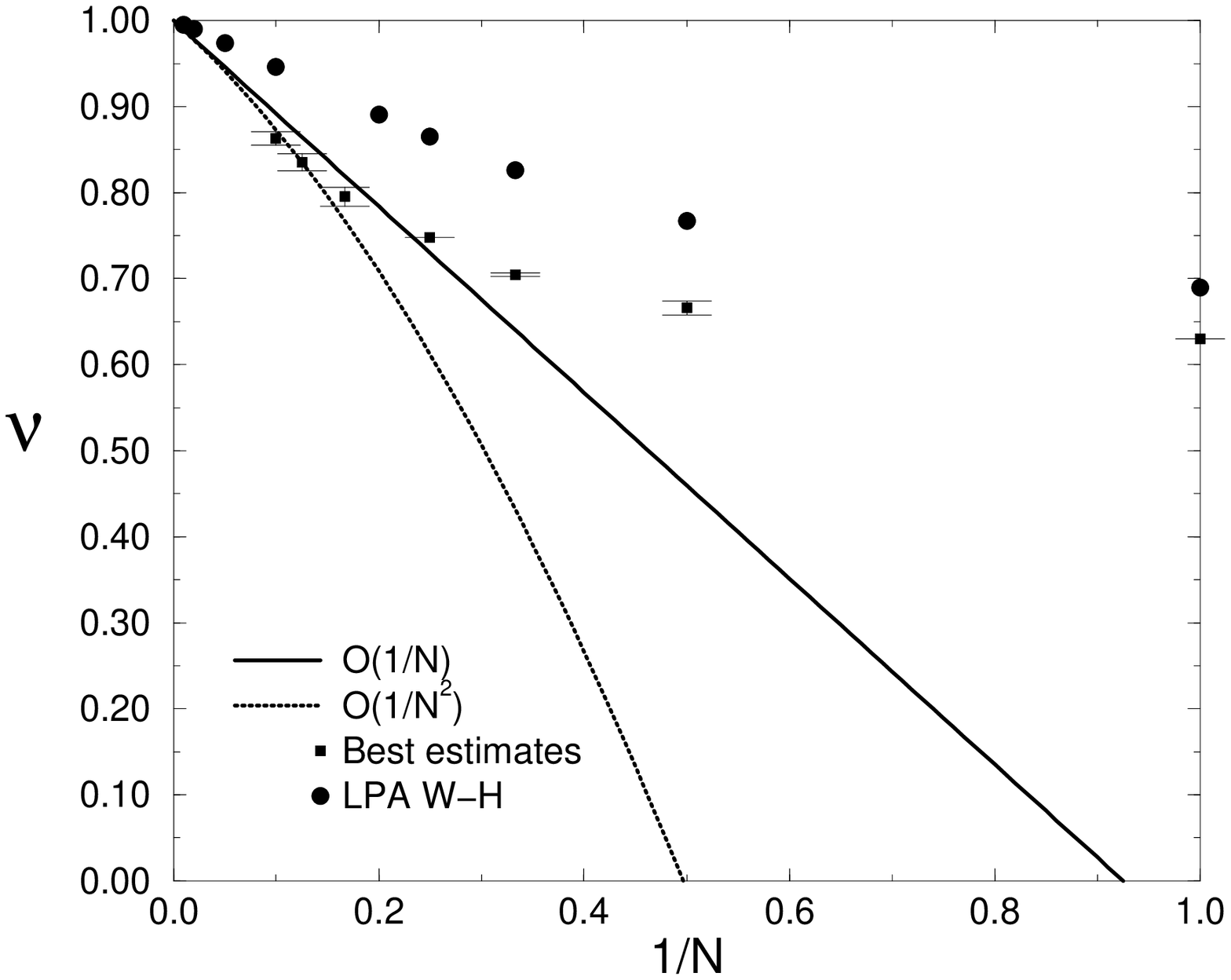}}
\parbox{6.5in}{
\footnotesize 
Figure 3: 
The critical exponent of the O($N$)-symmetric scalar theories in
$d=3$ as a function of $1/N$.
The results by the W-H equation in the LPA are shown with the
results of $O(1/N)$ and $O(1/N^2)$ in $1/N$-expansion.
The square points show the present best estimates summarized 
in Ref.\cite{BC}.
}
\vskip10mm
\epsfxsize=0.75\textwidth
\centerline{\epsffile{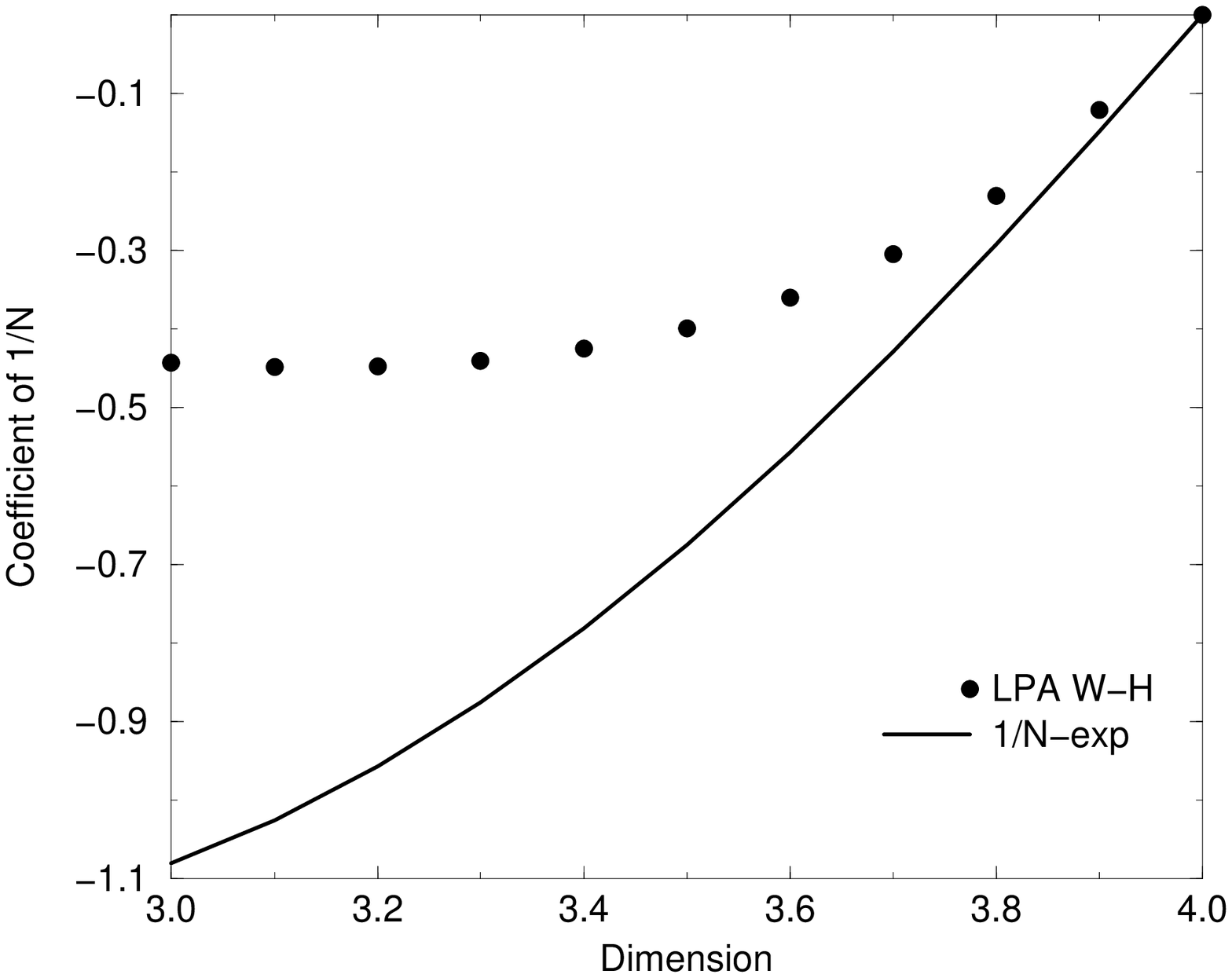}}
\parbox{6.5in}{
\footnotesize 
Figure 4: The dimensional dependence of the coefficient of $1/N$
in the W-H equation and in $1/N$-expansion. 
}
\end{figure}

Indeed the LPA seems to be inferior to $1/N$-expansion for
a large but finite $N$. This is because the estimation of $\nu$ by the RG
deviates considerably from the exact value in $O(1/N)$. 
Of course it is well expected that the next-to-leading 
calculation in the derivative expansion will become fairly 
accurate in this region. However such study has not been carried out yet. 
Here we numerically evaluate the coefficient of $O(1/N)$ for the
exponent $\nu$ calculated by the W-H equation for $3 \leq d \leq 4$. 
The results are shown in Fig.4 
with the coefficient of $1/N$ in Eq.(27).
The difference is getting smaller and smaller
as the dimension $d$ approaches to 4, because the effects of 
the derivative couplings decrease monotonically and 
eventually vanishes in  $d=4$ where the LPA is exact for any $N$.

\newpage

\section{Summary and discussions}

In this article it has been shown explicitly that the LPA of
the W-H equation becomes exact in $O(\epsilon)$ and in large
$N$ limit. This is because the higher order corrections involving
the derivative interactions are not generated in these order
calculations. The deviations appearing in the next orders in these
expansions are also studied. In Fig.5 the global behavior of the
exponents obtained by the W-H equation in the LPA as well as the
accurate ones are shown so as to summarize our numerical results.
It should be noted here that the RG method is rather effective even in 
the LPA over the whole parameter region of $N$ and $d$. Thus we can
comprehend the gross properties of the exponent by means
of our simple Wilson RG method.

\begin{figure}[htb]
\centerline{\epsfxsize=0.60\textwidth\epsffile{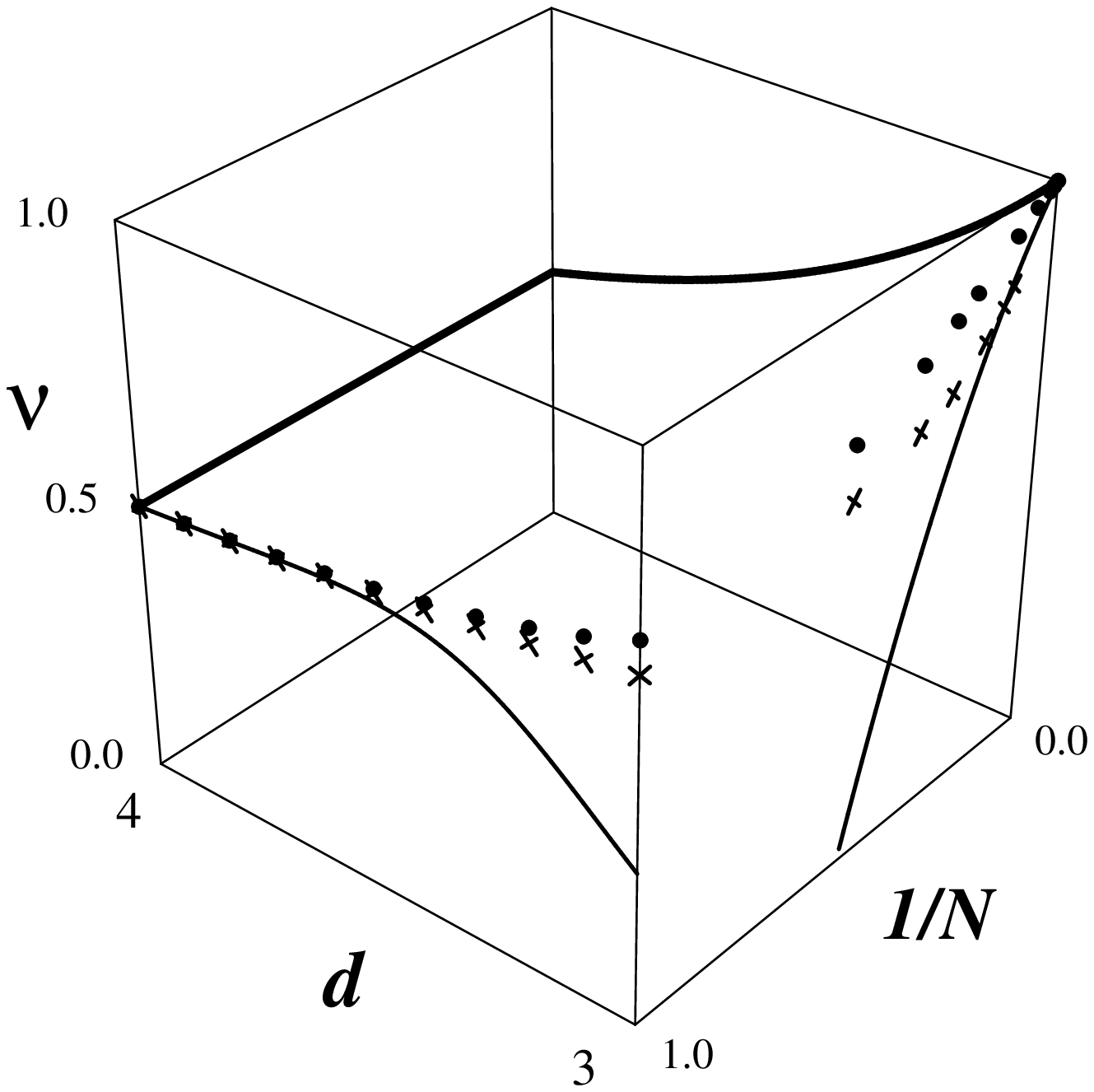}\hskip5mm
\vbox{\epsfxsize=0.25\textwidth\epsffile{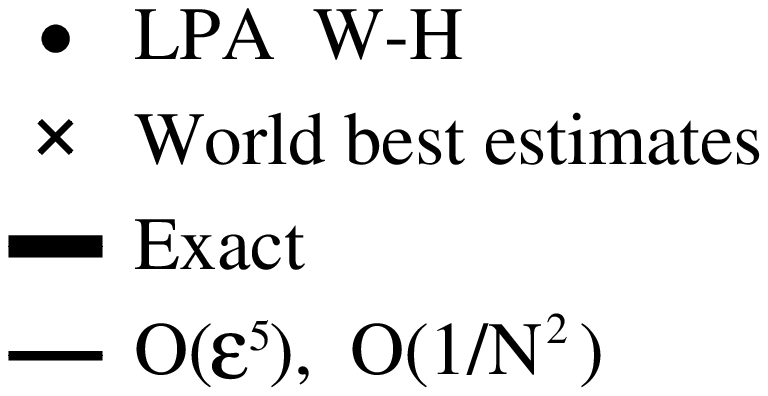}\vskip60mm}}
\vskip5mm
\parbox{6.5in}{
\footnotesize 
Figure 5: The variation of the exponent with respect to $N$ and $d$.
The fifth order of $\epsilon$-expansion and the second order of 
$1/N$-expansion are plotted for comparison. The world best estimates
are taken from Ref.\cite{BC}}
\end{figure}

The perturbative expansion as well as $\epsilon$-expansion leads to
the asymptotic series. Therefore such expansions cannot be applied
directly at the region where the expansion parameters are not small
enough. It is necessary to sum up these series through the Borel
transformation in order to obtain
quantitatively reliable results. Moreover we have to know the
large order behavior of the asymptotic series to carry out the
summation. In contrast to these expansion schemes, it is
remarkable that we can estimate the exponents by solving the 
W-H equation without resorting to the complicated manipulations 
of the resummation. 

Needless to say it is quite important to see whether
the higher order calculations in the derivative expansion generate
converging results or not. So far the Wilson RG equations have been
investigated up to the next-to-leading order of the derivative
expansion\cite{M2}.
By these analyses the exponents have been found to become
closer to the expected values in $O(\partial^2)$ than in the LPA.
It should be noted that the sharp cutoff scheme applied 
to derive the W-H equation 
leads to the difficulties of non-analyticity
when the derivative couplings are incorporated\cite{M3}. 
Therefore the formulations defined by the smooth cutoff\cite{P,W} have
been used in the next order calculations\cite{AMSST}.

The derivative expansion is not an ``expansion''.
It is just enlarging the functional subspace step by step in which the
Wilson RG equation is solved. Therefore it is much more like 
increasing the total lattice size in the Monte Carlo simulations. 
Nobody may expect any divergent behavior of the physical 
quantities when increasing the lattice size. 
Rather we should expect convergence, or oscillation at worst.
We have already met an example of this type of convergence in 
Fig.1, where we enlarge the dimension of the functional subspace
one by one and we get the strongly converging results 
(or oscillation, depending on the choice of the truncation scheme).
Thus our method of the Wilson RG equation may not suffer from
any divergent series.

Lastly let us discuss one of the applications to particle physics.
The perturbative beta functions have been known to show the behavior
of the asymptotic series as well. Therefore we cannot assert even
the existence of the non-trivial fixed point by using the perturbative
beta functions. Indeed the Borel summation technique has been
succeeded in two and three dimensions. In four dimensions, however,
even the Borel summability for the scalar theories has not been
clarified yet. Therefore the other non-perturbative methods are
desirable in four dimensions. For instance the argument of the 
triviality mass bound for Higgs boson and for top quark relies 
entirely on the non-existence of the non-trivial UV fixed point 
in the standard model.
Actually several works on the triviality bound by means of the Wilson
RG equations have been already done\cite{HN}. 
However more minute and realistic investigations would be desired
\cite{ASSTT}.

K-I.~A. is supported in part by the Grand-in-Aid for Scientific
Research (\#04640297) from the Ministry of Education.

\end{document}